\documentclass[twocolumn,superscriptaddress,showpacs,amsmath,amssymb,aps,prb]{revtex4-1}

\usepackage{graphicx}
\usepackage[pdftex, colorlinks=true, linkcolor=blue, citecolor=blue, urlcolor=blue]{hyperref}
\usepackage{times}

\DeclareGraphicsExtensions{.pdf}

\newcommand{\ioannina}{Department of Physics, University of Ioannina, Ioannina 45110, Greece}
\newcommand{\strasbourg}{Institut de Physique et Chimie des Mat\'{e}riaux de Strasbourg, 
                         Universit\'{e} de Strasbourg, CNRS UMR 7504, F-67034 Strasbourg, France}
\begin{document}
\title{Parity-dependent localization in $N$ strongly coupled chains}

\author{Dietmar Weinmann}
\affiliation{\strasbourg}
\author{S. N. Evangelou}
\affiliation{\ioannina}
\date{\today}

\begin{abstract}
Anderson localization of wave-functions at zero energy in quasi-$1D$ systems of $N$ disordered chains 
with inter-chain coupling $t$ is examined. Localization becomes weaker than for the $1D$ disordered chain 
($t=0$) when $t$ is smaller than the longitudinal hopping $t'=1$, and localization becomes usually much 
stronger when $t\gg t'$. This  is not so for all $N$. We find ``immunity" to strong localization for 
open (periodic) lateral boundary conditions when $N$ is odd (a multiple of four), with localization that 
is weaker than for $t=0$ and rather insensitive to $t$ when $t \gg t'$.
The peculiar $N$-dependence and a critical scaling with $N$ is explained by a perturbative treatment 
in $t'/t$, and the correspondence to a weakly disordered effective chain is shown. 
Our results could be relevant for experimental studies of localization in photonic waveguide 
arrays.  
\end{abstract}

\pacs{73.20.Fz, 71.55.Jv, 71.23.-k}
\maketitle

\section{Introduction}

Is everything known about Anderson localization\cite{1,2}? The answer is, surprisingly, 
negative. Since in real devices we cannot completely get rid of disorder, the subject, 
although over 56 years old, still remains an active part of condensed matter physics 
and from time to time holds surprises (see recent activity in topological insulators and 
superconductors).\cite{3}

Anderson localization is a phenomenon of complex quantum interference of electron 
waves in a disordered medium and also occurs for classical waves in disordered media. 
The extended states (ballistic or diffusive) of a metal turn into the exponentially 
localized states of an insulator which are restricted to a finite region of a disordered system. 
The extension of states is measured by their localization length $\xi$, 
which is infinite for extended states and finite for localized states. 
The states cease to be extended when the disorder $W$ exceeds some critical value $W_\mathrm{c}$. 
The main result of the so-called scaling theory of localization \cite{4} is that in the 
absence of symmetry breaking mechanisms (e.g.\ of time-reversal symmetry due to a 
magnetic field $\vec{B}$ or of spin rotation symmetry due to spin-orbit coupling SOC) 
a non-zero critical disorder ($W_\mathrm{c}>0$) exists for the localization of all states in $3D$, 
while the states become localized (have non-infinite $\xi$'s) 
even for infinitesimally small disorder ($W_\mathrm{c}=0$) in $1D$ and $2D$.\cite{5,6} 

The subject is still active both experimentally and theoretically for at least three reasons: 
first, there is a large variety of situations that depend on the kinds of disorder and the 
symmetry of the Hamiltonian, e.g. the 10 symmetry classes of localization, \cite{7} and the 
crossover behavior between them. 
Those symmetry classes are the three basic, orthogonal, unitary (in the presence of magnetic 
field $\vec{B}$), and symplectic (systems with spin-orbit-coupling (SOC)), the 3 chiral, and the 
4 Bogolyubov-de Gennes classes for superconducting systems. 
A second reason is the emergence of topological structures \cite{3} in which some states 
are immune to Anderson localization, at least for weak disorder. They occur via 
spectral gaps which support protected states 
(e.g. in $2D$, conductance quantization in the presence of $\vec{B}$ for the quantum Hall 
effect and in the presence of SOC for the quantum spin Hall effect). For example, off-diagonal 
disorder with random hopping belongs to one of the 3 chiral classes and an even-odd 
effect was observed at zero energy for $N$ coupled chains, with a diverging density of 
states $\rho(E)$ and a localization length that depends on the parity of $N$.\cite{8,9} 
A final reason for the subject being active today is that wave phenomena in disordered 
media can also be investigated for electromagnetic waves in complex structures, such as 
waveguide arrays which resemble a finite lattice. A wealth of recent experimental results 
\cite{10,11,12,13} combine disorder with non-linearity, and it is possible to test various 
theories (e.g. of topological effects, the critical exponents at the Anderson transition, 
etc.), including regimes that are difficult to access in disordered electronic 
systems.

\begin{figure}
\centerline{\includegraphics[width=\linewidth]{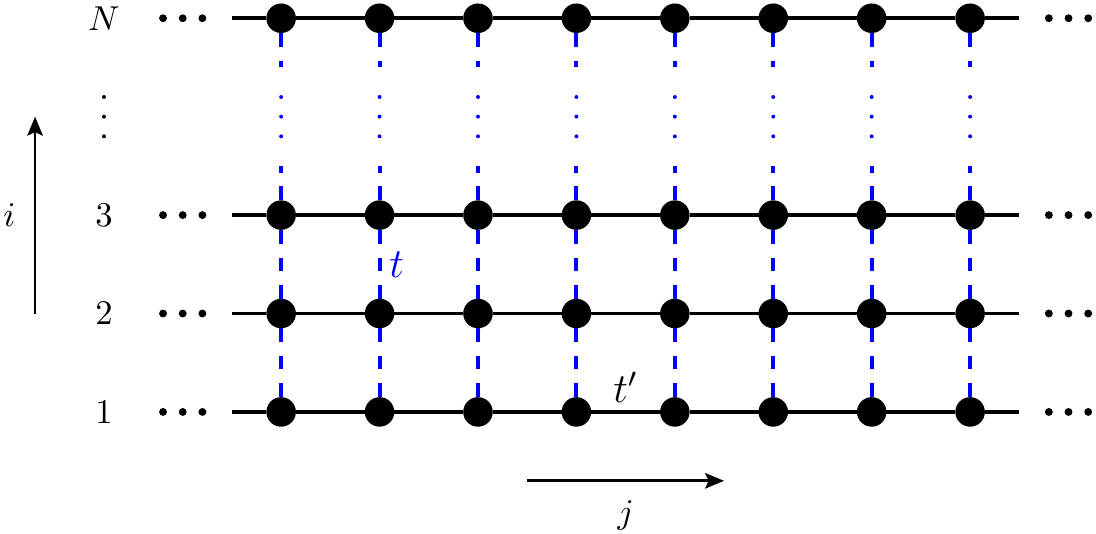}}
\caption{\label{fig:sketch}
The quasi-1D system of $N$ disordered chains, with on-site disorder of strength $W$, 
longitudinal hopping $t'=1$ (black solid lines) 
and inter-chain hopping $t$ (blue dashed lines).}
\end{figure}
Anderson localization in a disordered chain ($1D$) is very different from that in a disordered 
plane ($2D$) made of $N$ coupled disordered chains with $N\to \infty$. 
The present work focuses on an assembly of $N$ coupled $1D$ chains (see Fig.\ \ref{fig:sketch}), 
which is a quasi-$1D$ disordered system of increasing width ($N$ is the number of 
its rows), that is expected to tend to a $2D$ lattice as $N$ increases. 
The crossover from $1D$ to $2D$ is studied in the presence of diagonal disorder ($W$) 
and as a function of the inter-chain coupling of strength $t$ (the longitudinal hopping is $t'=1$). 
 
Our main result, from a study for a wide range of values of the ratio $t/t'$ and various $N$, 
is an unexpected $t$-dependence for special $N$'s, 
where localization for the quasi-$1D$ system remains weak in the 
strong-coupling limit $t\gg t'$ (even weaker than at $t=0$), and rather insensitive to $t$. 
For certain $N$'s Anderson localization is more pronounced as $t$ increases and not for other 
$N$'s where there is ``immunity" to strong localization. This ``immunity" occurs when $N$ is odd 
for open lateral boundary conditions (BC) and when $N$ is a multiple of four for periodic lateral BC. 
In other words, in the quasi-$1D$ disordered system with strong inter-chain coupling $t$, 
Anderson localization is weakened by the large inter-chain coupling only for these special $N$'s. 
For other $N$'s localization becomes stronger as $t$ increases. 

In Sec.\ \ref{sec:model} we present the model system of $N$ coupled disordered chains. 
The method of study and our results for the Lyapunov exponents can be found in 
Sec.\ \ref{sec:method}. The finite size scaling analysis of the results is presented in 
Sec.\ \ref{sec:scaling}, and their explanation in terms of a perturbative approach at 
strong inter-chain coupling is given in Sec.\ \ref{sec:perturbation}. 
We discuss the results in Sec.\ \ref{sec:discussion} before 
we present our conclusions in Sec.\ \ref{sec:conclusions}.

\section{Numerical approach to localization}
\label{sec:numerics}

\subsection{Quasi-1D model of $N$ coupled chains}
\label{sec:model}

We study Anderson localization of non-interacting particles in the phase-coherent quantum 
system of $N$ parallel disordered chains, sketched in Fig.\ \ref{fig:sketch}, with on-site 
disorder of strength $W$ and intra-chain hopping $t'=1$ which represents the energy scale. 
The inter-chain hopping $t$ takes a broad range of values, from very small ($t\ll t'$) to 
very large ($t\gg t'$). The system is a quasi-1D strip  of a square lattice having width $N$, 
on-site disorder $W$ and anisotropic hopping, namely $t'=1$ in the longitudinal direction 
and $t$ in the transverse direction, respectively. 
Labelling the sites by $\{i,j\}$ with $i=\{1,2,\dots,N\}$ the transverse and 
$j$ 
the longitudinal coordinate, the Hamiltonian reads
\begin{eqnarray}
H &=&\sum_{j}\left\{ \sum_{i=1}^{N} \left(\epsilon_{i,j} c^+_{i,j}c^{\phantom{+}}_{i,j} 
+ t'\left[ c^+_{i,j}c^{\phantom{+}}_{i,j-1}+\text{hc}\right]\right) \right. \nonumber \\
 && \phantom{\sum}\left. + t \left[\sum_{i=2}^{N} c^+_{i,j}c^{\phantom{+}}_{i-1,j} 
 + \eta c^+_{1,j}c^{\phantom{+}}_{N,j}+\text{hc}\right]\right\} \, ,
\label{eq:hamiltonian}
\end{eqnarray}
where $c^+_{i,j}$ ($c^{\phantom{+}}_{i,j}$) creates (annihilates) a particle on site 
$\{i,j\}$. The on-site energies $\epsilon_{i,j}$ are independent random variables drawn 
from a uniform distribution within $[-W/2;W/2]$. 
The parameter $\eta$ takes the values 0 and 1 for open and periodic lateral BC, 
respectively. 

We study localization in the full $t$-range, and increase $N$ to approach a 
$2D$ disordered system \cite{13} with anisotropy ($t\neq t'$) or without ($t=t'$). 
The single chain case is related to the uncoupled limit $t=0$ where an ensemble of 
independent chains exists. The ladders with $N=2$ and $N=3$ legs were extensively 
studied,\cite{14,15,16,17} also in the presence of magnetic and electric fields.

\begin{figure}
\centerline{\includegraphics[width=\linewidth]{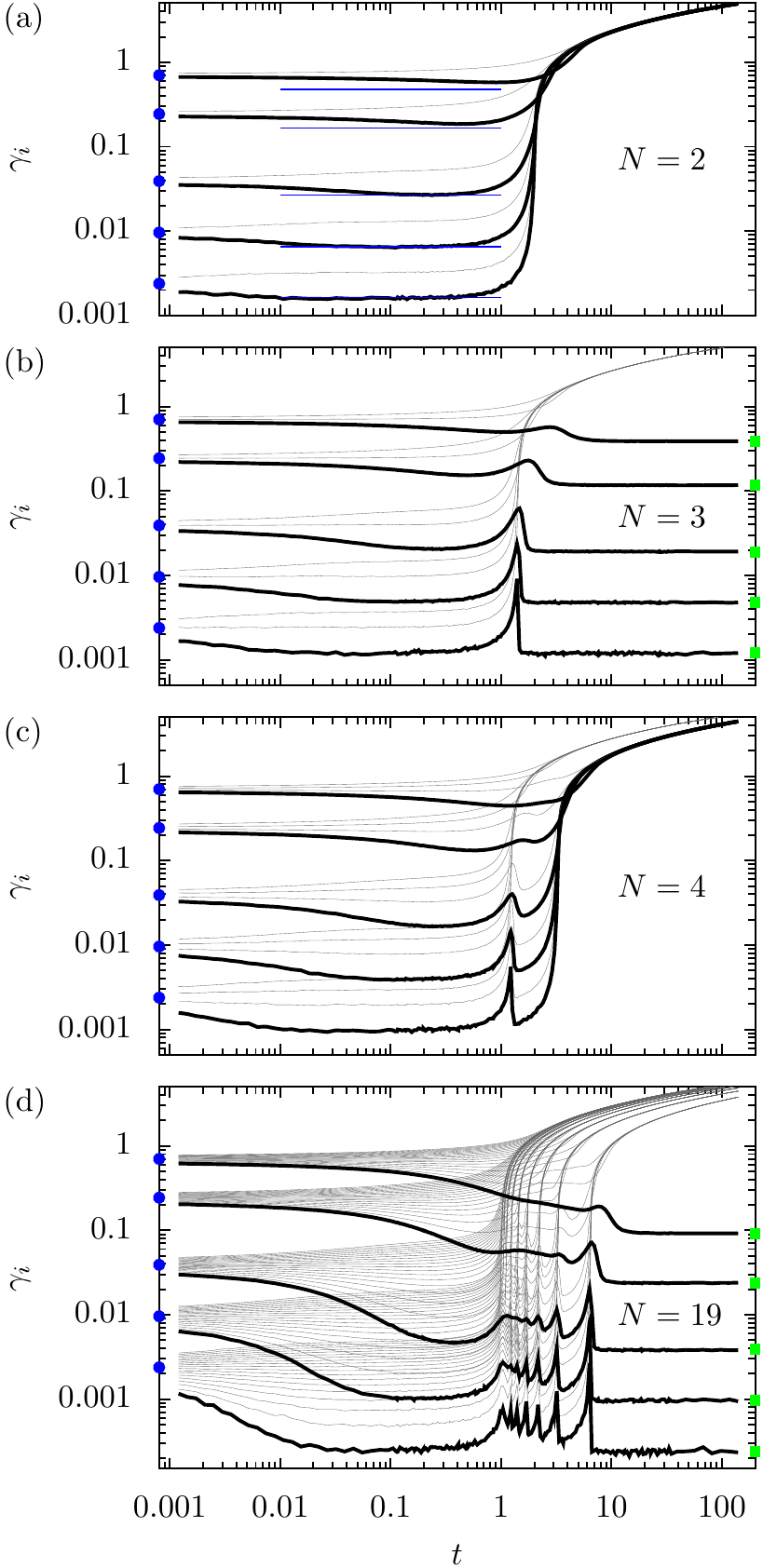}}
\caption{\label{fig:nbc}
(Color online)
The Lyapunov exponents $\gamma_{1},\gamma_{2},\dots ,\gamma_{N}$ at $E=0$
\textit{vs}. the inter-chain hopping $t$ for quasi-$1D$ disordered systems with open lateral 
BC and disorder strengths $W$=\{0.5, 1, 2, 5, 10\} from bottom up, with
(a) $N=2$, (b) $N=3$, (c) $N=4$, (d) $N=19$. 
For each disorder strength, the smallest exponent $\gamma_{1}$ is represented 
by a thick solid line. 
For odd $N=3$ and $N=19$, $\gamma_{1}$ is rather insensitive to $t$, for all $W$.
The limiting values at $t=0$ ($t/t'\to\infty$) are indicated by blue dots (green squares) 
on the left (right) axis. 
The horizontal blue solid lines in (a) indicate the values of Eq.\ \eqref{eq:dorokhov}.}
\end{figure}
\subsection{Method and results}
\label{sec:method}

The exponential dependence of the wave-functions along a disordered quasi-1D strip 
containing $N$ coupled chains (described in Sec.\ \ref{sec:model} and sketched in 
Fig.\ \ref{fig:sketch}) is extracted from numerically calculated products of $N\times N$ 
transfer matrices and characterized by $N$ Lyapunov exponents. \cite{5,6} 
We compute the exponents via the standard method of Gram-Schmidt reorthogonalization after 
about ten steps of matrix multiplication until convergence is reached for a very large number 
of steps along the chain. 
The adopted procedure corresponds to the factorization of the transfer matrix into a product 
of an orthogonal matrix and a diagonal matrix with positive matrix elements. 
The $N$ positive Lyapunov exponents $\gamma_{i}$ ($i=\{1,2,\dots ,N\}$) are subsequently defined 
as the positive logarithms of the elements of the diagonal matrix.\cite{18}

For a single disordered chain ($N=1$), the Lyapunov exponent $\gamma$ is the inverse of the 
localization length $\xi$ which describes the length scale of the exponential increase or 
decrease of the wave-function along the chain. 
A perturbative approach for small disorder $W$ at $E=0$ yields
\begin{equation}\label{eq:lyapunov1d} 
\gamma\simeq 0.0095\, W^{2}+O\left[W^{4}\right]
\end{equation}
for the $1D$ Lyapunov exponent. \cite{19}  

For the quasi-$1D$ system of width $N$, the exponential decay of the wave-functions with the 
longitudinal coordinate is determined by $N$ Lyapunov exponents $\gamma_{i}$. 
The smallest positive exponent $\gamma_{1}$ dominates the transport properties and determines 
the localization length $\xi=1/\gamma_1$ of the quasi-1D strip. 
The scaling behavior of $\gamma_{1}$ as a function of $N$ with an extrapolation to large 
$N\to \infty$ can be studied via the so-called finite size scaling technique \cite{5,6} which 
allows to determine the localization length of the infinite $2D$ system. 

In Fig.\ \ref{fig:nbc} we plot $\gamma_{i}$ at energy $E=0$ \textit{vs}. the inter-chain coupling $t$ 
over a wide range of values of $t$, for quasi-$1D$ strips with $N=2$, 3, 4, and 19 chains having 
open lateral BC. In each case, we show results for five disorder strengths between $W=0.5$ and $W=10$. 
The single chain ($N=1$) exponent $\gamma$ is indicated by blue circles on the left 
vertical axis. The five circles with increasing $\gamma$ correspond to the chosen values of $W$. 
While all $\gamma_{i}$'s are expected to converge to them in the limit $t\to 0$, a finite $t$ gives 
a spreading of the $\gamma_i$ for a given $W$. 
We focus on the smallest exponent $\gamma_{1}$ (solid lines) which gives the localization length 
$1/\gamma_{1}$ of the quasi-$1D$ system. 

\begin{figure}
\centerline{\includegraphics[width=\linewidth]{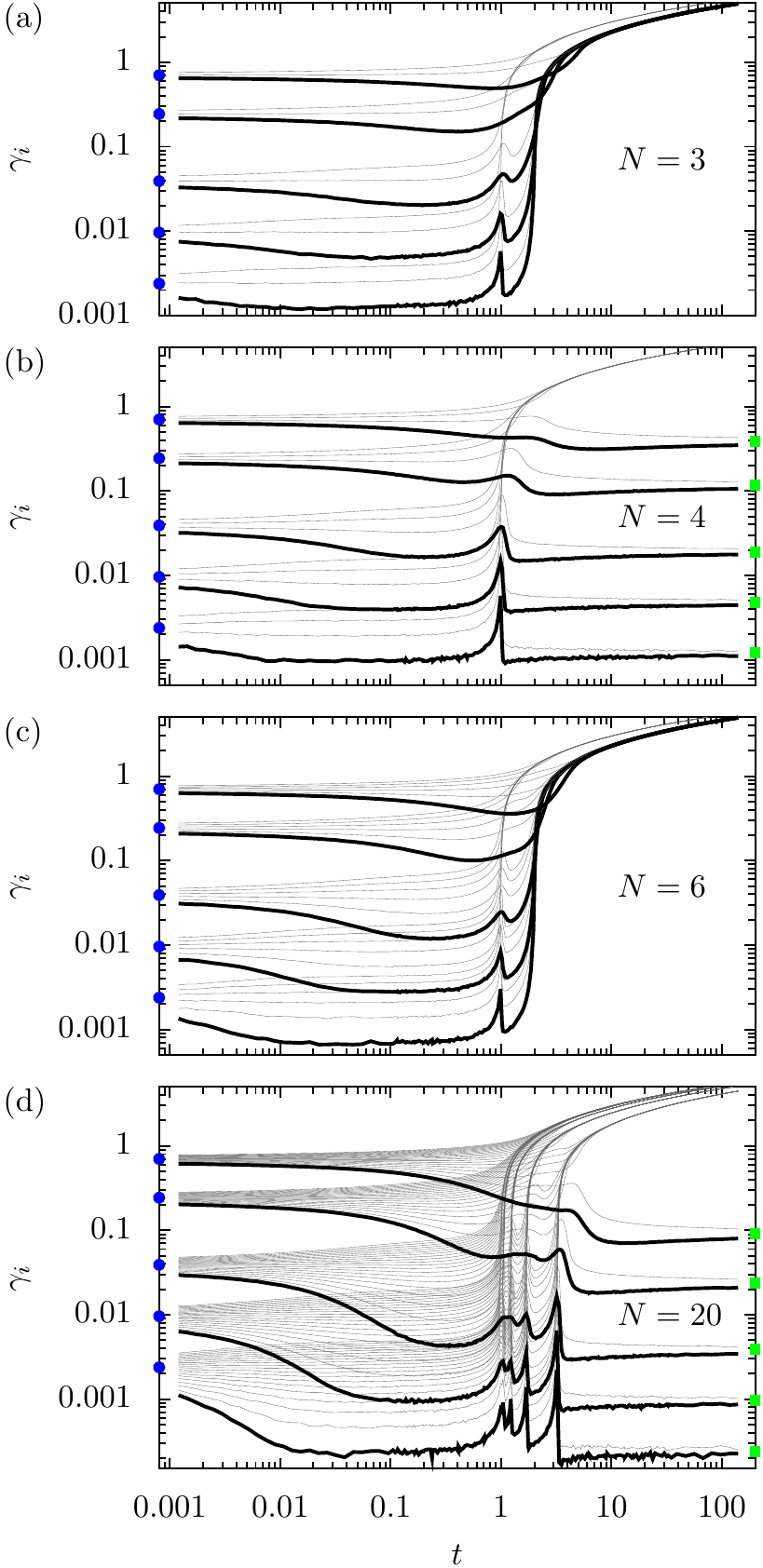}}
\caption{\label{fig:pbc}
(Color online) The Lyapunov exponents $\gamma_{1},\gamma_{2},\dots , \gamma_{N}$ \textit{vs}. 
the inter-chain hopping $t$ for quasi-$1D$ disordered systems with lateral periodic BC for 
disorder strengths $W$=\{0.5, 1, 2, 5, 10\} from bottom up, 
(a)  $N=3$, (b) $N=4$, (c) $N=6$, (d) $N=20$. For $N=4$ and $N=20$, multiples of four, 
the two smallest exponents $\gamma_{1}$, $\gamma_{2}$  
(thick lines) are rather insensitive to $t$.}
\end{figure}
The behavior of $\gamma_1$ at small $t$ is seen to be qualitatively similar for all $N$. 
The inter-chain coupling reduces localization and pushes $\gamma_1$ towards a minimum for 
$t\lesssim 1$. 
The horizontal blue lines in Fig.\ \ref{fig:nbc} (a) indicate Dorokhov's perturbative 
prediction \cite{14} for the minimum value of $\gamma_1$  
\begin{equation}
\label{eq:dorokhov}
\gamma_{1}^\mathrm{min}\approx \left(1-\frac{1}{\pi}\right)\gamma \, ,
\end{equation} 
obtained for a ladder of $N=2$ chains with coupling $t\leq 1$ ($\gamma$ is the Lyapunov 
exponent for a single chain $N=1$). 
The minimum value of Eq.\ \eqref{eq:dorokhov} has the same $W$-dependence as $\gamma$,
and the perturbative value \eqref{eq:lyapunov1d} becomes less accurate for larger $W$. 
While weakly coupled chains are well understood, we now focus our interest on the case of 
strong coupling.

In the regime of strong inter-chain coupling ($t$ large), striking differences between different $N$'s 
are observed. For even $N$ (see  Figs.\ \ref{fig:nbc} (a) and (c)) all Lyapunov exponents of the quasi-$1D$ 
system, including the smallest exponent $\gamma_1$, increase to very large values. This implies 
increasingly strong localization, with a vanishing localization length when $t \to \infty$. 
In contrast, for odd $N$'s (Figs.\ \ref{fig:nbc} (b) and (d)), only the $N-1$ largest  
exponents increase with $t \to \infty$ while the smallest exponent $\gamma_1$ assumes a 
small value that depends on $W$. 
In the strong coupling regime, $\gamma_1$ is almost independent of $t$ and its asymptotic values in 
the limit $t\to \infty$ (green squares on the right vertical axis) lie below the single chain result 
(blue circles on the left vertical axis). Between the regimes of weak and strong coupling, $\gamma_1$ 
goes through one or more maxima at moderate coupling strength $t\gtrsim 1$.  

In Fig.\ \ref{fig:pbc} we show the Lyapunov exponents calculated for quasi-1D strips 
with periodic lateral BC, at the same energy and disorder values as in Fig.\ \ref{fig:nbc}. 
The results for $N=$ 3, 4, 6, and 20 show the same qualitative behavior at small $t$, as for open BC. 
In the regime of strong inter-chain coupling $t$, the behavior is different. 
For periodic BC localization strongly increases with large $t$ whenever 
$N$ is not a multiple of four, very much like in the case of even $N$ and open BC (Fig.\ \ref{fig:nbc}). 
For the values of $N$ that are multiples of four, the \textit{two} smallest Lyapunov 
exponents $\gamma_{1}$ and $\gamma_{2}$ display the peculiar behavior of very weak dependence on 
$t$ and assume very low values at large $t\to \infty$. 
The smallest exponent $\gamma_1$ approaches the strong coupling limit 
(green squares on the right vertical axis) from below, and has its minimum value 
at moderately strong $t$, just above the region of the peaks observed at $t \gtrsim 1$. 

The number of peaks in the $\gamma_1$'s for $t\gtrsim 1$ increases with $N$. 
More peaks appear for open BC (Fig.\ \ref{fig:nbc}) than for periodic BC (Fig.\ \ref{fig:pbc}). 
The small $\gamma_1$ for certain $N$ (in Figs.\ \ref{fig:nbc} and \ref{fig:pbc}) 
and the very weak dependence on $t$ for $t\gg t'$ is in striking contrast with the strong increase 
of localization occurring for other $N$'s. 

\subsection{Finite size scaling}
\label{sec:scaling}

\begin{figure}
\centerline{\includegraphics[width=\linewidth]{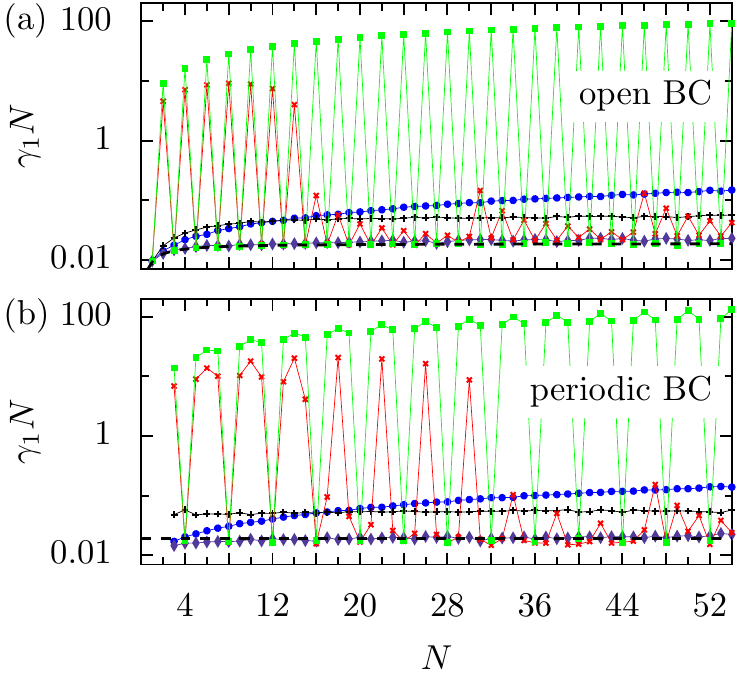}}
\caption{\label{fig:gamma_N}
(Color online) The scaling of $\gamma_{1}N$  {\textit {vs.}} the width $N$ for a quasi-$1D$ strip having 
(a) open lateral BC and (b) periodic BC, at $E=0$ and $W=1$. 
Blue circles are for $t=0.01$, violet diamonds for $t=0.1$, black pluses for $t=1$, 
red crosses for $t=10$, and green squares for $t=100$. Lines are guides to the eye. 
The oscillations of $\gamma_1$ as a function of $N$ with period two (four) 
for open (periodic) BC emerge with increasing $t$ when $t>t'$ (red crosses and green squares).
Moreover, the low values of $\gamma_{1}N\approx 0.01$ at the minima of those oscillations are 
approximately independent of $N$.
The dashed lines in the lower parts of the figure are the perturbative results for 
the strong coupling regime of Sec.\ \ref{sec:perturbation}.}
\end{figure}
In order to extract the behavior at large $N$ and see the transition to $2D$, we studied the 
scaling properties of $\gamma_{1}$ with the number of chains $N$ (the  width of the quasi-1D strips). 
The finite size scaling theory predicts that an increase of $\gamma_1 N$ with $N$ points towards 
localization in $2D$, a decrease of $\gamma_1 N$ with $N$ indicates delocalization, and  a constant 
value corresponds to intermediate (critical) behavior. In the localized case, the product $\gamma_1 N$ 
increases proportional to $N$ and the constant of proportionality gives the inverse 
localization length of the $2D$ system.\cite{5,6}

In Fig.\ \ref{fig:gamma_N} we plot $\gamma_1 N$ \textit{vs}. $N$ in a semi-log plot for $W=1$, 
$E=0$ and different values of $t$. 
For very small $t$ (blue circles) and $t\approx 1$ (black pluses) a smooth $N$-dependence is observed, 
in sharp contrast with the $N$-dependent oscillations found for large $t$ (red crosses) and very 
large $t$ (green squares). These oscillations correspond to the peculiar behavior of localization in 
the strong coupling regime already shown in Figs.\ \ref{fig:nbc} and \ref{fig:pbc}.
In Fig.\ \ref{fig:gamma_N} (a) and (b), the period of two for open BC and four for periodic BC is seen, 
respectively. The maxima of $\gamma_{1}$ indicate very strong localization while much weaker 
localization occurs for open (periodic) BC when $N$ is odd (multiple of four), even weaker than for 
small coupling ($t\ll 1$).

In Fig.\ \ref{fig:gamma_N}, for very small $t=0.01$ (blue circles) $\gamma_{1}$ is close 
to the single chain value and $\gamma_1 N$ increases with $N$ indicating localization in the 
large-$N$ limit. A similar increase (at higher values of $\gamma_1 N \gtrsim 10$) is observed 
for the maxima of the $N$-dependent oscillations (green squares).
In contrast, only a very weak increase can be detected for moderately weak inter-chain coupling 
$t=0.1$ (violet diamonds), and for the isotropic case $t = 1$ (black pluses) the $\gamma_1 N$ appear 
to be independent of $N$ up to the largest $N$ we considered. 
The same qualitative behavior, with reduced Lyapunov exponents, is shown for the minima at 
strong coupling. Their values are close to the ones obtained for moderate inter-chain coupling 
(violet diamonds).  

A particular behavior is seen for rather large $t=10$ 
(red crosses in Figs.\ \ref{fig:gamma_N} (a) and (b)). The $N$-dependent oscillations disappear 
for large $N$, where the minima remain low but the maxima are considerably reduced and tend 
towards the low minimum values. Other values of disorder gave qualitatively similar behavior, 
with increasing Lyapunov exponents when increasing the disorder. For stronger disorder $W$ 
(Figs.\ \ref{fig:gamma_N} (a), (b) is for $W=1$), larger values of $t$ are needed to make the 
$N$-dependent oscillations appear. 
This leads to the conjecture that oscillations with $N$ can be observed in the strong 
coupling regime for $t\gtrsim N$ or $t\gtrsim W$, while localization is weak for all $N$ in an 
intermediate regime with $N \gtrsim t \gtrsim 1$.

The intermediate critical-like scaling 
\begin{equation}\label{eq:scaling}
\gamma_{1} N \propto \mathrm{const.}\, 
\end{equation} 
is observed for the isotropic system ($t=t'$) for not very large $N$ (the localization length in $2D$ 
is huge), and for the minima of the $N$-dependent oscillations at strong coupling $t$. 
The scaling \eqref{eq:scaling} was first suggested by Thouless \cite{20} and is known to occur 
for the $E=0$ state in quasi-$1D$ disordered systems with off-diagonal disorder \cite{21}, 
in Carbon nanotubes \cite{22}, etc.  

In Fig.\ \ref{fig:gamma_N}, however, $W=1$ which is rather small and the corresponding $1D$ localization 
length at $E=0$ (see \eqref{eq:lyapunov1d}) is {$1/\gamma \simeq 104$}, larger than the 
largest $N$ considered in our calculations. The obtained law is most probably a crossover regime 
which can turn into a localized scaling (the behavior of Eq.\ \eqref{eq:scaling} changes to 
$\gamma_{1} N \propto N$) when $N$ becomes so large that the lateral extension of the 
states is limited by localization rather than the system width $N$.
Since the $2D$ localization length is larger than in $1D$ and in $1D$ proportional to the 
square of the hopping element, the critical scaling \eqref{eq:scaling} is expected to hold in 
a crossover region that extends to sizes of at least $N\sim 104 (t/W)^2$. 
When the transverse coupling $t$ increases, the critical scaling region can become very large. 
However, numerical investigation of increasing values of $N$ requires 
increasing computer power making it difficult to observe the transition in scaling behavior 
for $W=1$. We have checked that for strong disorder $W=10$ (when the $1D$ localization length 
$1/\gamma \simeq 1$ is small), $\gamma_{1}N$ increases with $N$ in all cases, 
indicating localization in $2D$ and supporting the scenario discussed above. 

\section{Perturbation theory for the strong coupling regime}
\label{sec:perturbation}

The intriguing $N$-dependent oscillations found numerically (see Sec.\ \ref{sec:numerics}) and the 
weakness of localization in the strong coupling regime for special values of $N$ motivate us to seek 
an analytical understanding of the localization behavior at large inter-chain coupling $t$.
We have developed a theory that is appropriate when the last term of the Hamiltonian \eqref{eq:hamiltonian}, 
containing the inter-chain coupling $t$, dominates over the disorder $W$ and the intra-chain hopping $t'$. 
The parameters $t'/t, W/t$ can then be treated as small perturbations. 

The starting point is the strong coupling limit $t\to \infty$ which corresponds to the unperturbed case 
$t'/t=W/t=0$. In this limit, the longitudinal coupling is negligible and the quasi-$1D$ system consists 
of uncoupled transverse slices composed of $N$ sites.
We show in the sequel that taking into account $t'/t$ and $W/t$ in lowest order allows, 
for special values of $N$, to map the perturbed quasi-$1D$ system onto an effective weakly disordered 
$1D$ chain. 

In the limit $t'/t=0$ the $N$ disordered chains become uncoupled and the Hamiltonian reduces to a sum $H=\sum_j H_j$ of independent blocks $H_j$, each of them describing a transverse slice at the longitudinal position that is given by the  index $j$.
The eigenenergies and the eigenstates of each slice can be obtained by diagonalizing the corresponding block $H_{j}$ of the Hamiltonian. 

\subsection{Open boundary conditions}

As an example, for $N=3$ and open BC, the Hamiltonian for the $j$-th slice is
\begin{eqnarray}
H_j &=& \left( \begin{array}{ccc}
\epsilon_{1,j} & t & 0 \\
t & \epsilon_{2,j} & t \\
0 & t & \epsilon_{3,j} 
\end{array} \right)
\\
&=& t \left( \begin{array}{ccc}
\epsilon_{1,j}/t & 1 & 0 \\
1 & \epsilon_{2,j}/t & 1 \\
0 & 1 & \epsilon_{3,j}/t 
\end{array} \right) \, .
\end{eqnarray}
In the limit $W/t\to 0$, the diagonal elements can be neglected as compared to the non-zero off-diagonal 
elements, and one has in zeroth order the Hamiltonian  
\begin{equation}
H_j^{(0)} = t \left( \begin{array}{ccc}
0 & 1 & 0 \\
1 & 0 & 1 \\
0 & 1 & 0 
\end{array} \right)\, ,
\end{equation}
which is independent of $j$.
The eigenvalues of $H_j^{(0)}$ are 
\begin{equation}
E_{1}^{(0)}=0 \quad \text{and} \quad E_{2/3}^{(0)}=\pm t\sqrt{2}\, .
\end{equation}
The two eigenvalues $E_{2/3}^{(0)}$ disappear to $\pm \infty$ in the strong coupling limit and are 
thus irrelevant for the behavior of the system at finite energy. In contrast, the first eigenvalue 
$E_{1}^{(0)}$ is independent of $t$ and crucial for the $E=0$ properties. 
The eigenstate of $H_j^{(0)}$ that corresponds to $E_{1}^{(0)}=0$ is
\begin{equation}\label{eq:slicestate}
\left|\psi_{1,j}^{(0)}\right\rangle = \frac{1}{\sqrt{2}} \left(c^{+}_{1,j}-c^{+}_{3,j}\right)|0\rangle\, ,
\end{equation}
where $|0\rangle $ is the vacuum state.
There is one such state for each value of the longitudinal index $j$, reducing the $N$ sites of the slice $j$ to a single relevant level.    

We now consider the first-order corrections in $W/t$ and $t'/t$ that lead to an effective disorder 
and a coupling of the slices to obtain an effective chain along the longitudinal ($j$) direction. 
In lowest order in $W/t$, the energies of the slice levels \eqref{eq:slicestate} are determined 
by the random energies $|\epsilon_{i,j}|\ll t$ of the original system as
\begin{equation}\label{eq:e1}
E^{(1)}_{1,j}=\left\langle \psi_{1,j}^{(0)}\right|H\left|\psi_{1,j}^{(0)}\right\rangle 
= \frac{\epsilon_{1,j} + \epsilon_{3,j}}{2} \, .
\end{equation}
A small $t'/t$ couples the slices, with hopping matrix elements in lowest order given by 
\begin{equation}
\left\langle \psi_{1,j\pm 1}^{(0)}\right|H\left|\psi_{1,j}^{(0)} \right\rangle = t' \, .
\end{equation}
The effective $1D$ chain which is obtained for strong coupling therefore has hopping $t'$ 
and random energies given by the slice energies $E^{(1)}_{1,j}$ of Eq.\ \eqref{eq:e1}.
Its disorder is reduced as compared to the disorder of the original quasi-$1D$ model because 
the on-slice energies \eqref{eq:e1} are averages of the independent on-site random energies 
$\epsilon_{1,j}$ and $\epsilon_{3,j}$. The localization along the effective chain is therefore 
weaker than that of a single chain in the quasi-$1D$ geometry. 
The on-slice energies $E^{(1)}_{1,j}$ have no longer uniform probability distributions and 
the disorder variance for $N=3$ is reduced by a factor of two with respect to that of the 
original quasi-1D strip. This mechanism is responsible for the reduction of localization 
observed at $E=0$ for strong coupling $t$ (see Fig.\ \ref{fig:nbc}).

The perturbative approach is easily generalized to arbitrary values of $N$ and also 
to periodic lateral boundary conditions. For open boundary conditions and odd $N$, 
the clean slices always have one eigenenergy $E_{1}^{(0)}=0$, which gives rise to an 
effective $1D$ chain. 
For even $N$ all slice eigenenergies tend to $\pm\infty$ in the strong coupling limit 
$t\to\infty$ and no effective chain exists at $E=0$. 
This readily explains the even-odd oscillations for the localization strength observed 
in the strong coupling limit. 

For odd $N$ an energy $E_{1}^{(0)}=0$ is found for the unperturbed slices and the effective 
chain appearing in lowest order in $t'/t$ and $W/t$ has hopping elements equal to $t'$ 
(independent of $N$), and on-slice energies 
\begin{equation}
E^{(1)}_{1,j}=\frac{2}{N+1}\left(\epsilon_{1,j} + \epsilon_{3,j} 
                     + \epsilon_{5,j} + \dots + \epsilon_{N,j}\right) \, .
\end{equation}
The calculated Lyapunov exponents for those effective chains with different $W$ (shown 
in Figs.\ \ref{fig:nbc} (b) and \ref{fig:nbc} (d) as green squares on the right vertical axis) are 
the asymptotic values assumed by the lowest exponents $\gamma_1$ of the quasi-$1D$ strips in the strong 
coupling limit.
 
The variance of the probability distribution of the on-slice energies $E^{(1)}_{1,j}$ is an average 
over $(N+1)/2$ of the on-site energies $\epsilon_{i,j}$ and given by $\sigma^2=2 \sigma^2_0/(N+1)$. 
It is smaller than the variance $\sigma^2_0 = W^2/12$ of the uniform distribution within 
$[-W/2;W/2]$ (from which the $\epsilon_{i,j}$ are drawn).
The same variance is obtained from a uniform distribution which has the reduced effective 
disorder strength 
\begin{equation}\label{eq:disorder-eff}
W_\mathrm{eff}=W\sqrt{\frac{2}{N+1}} \, .
\end{equation}
At $E=0$ and for weak disorder, the effective $1D$ chain 
with open BC and odd $N$ obeys Eq.\ \eqref{eq:lyapunov1d}. Using the effective disorder $W_\mathrm{eff}$
\eqref{eq:disorder-eff}, one gets the Lyapunov exponent 
\begin{equation}\label{eq:lyapunov-eff-nbc}
\gamma_\mathrm{eff}\simeq \frac{0.019}{N+1} W^2 + O\left[W^4\right]\, .  
\end{equation}
The values of $\gamma_\mathrm{eff}$ (shown by a dashed line in Fig.\ \ref{fig:gamma_N} (a))
are in excellent agreement with the numerical data for strong $t$. 

\subsection{Periodic boundary conditions}

In the case of periodic BC the clean slice has two degenerate eigenenergies
$E_{1/2}^{(0)}=0$ when $N$ is a multiple of four. Otherwise, all eigenenergies of the 
slice are proportional to $t$ and go to $\pm \infty$ in the strong coupling limit.
The space of the zero energy eigenstates is spanned by the degenerate states
\begin{eqnarray}\label{eq:PBC-psi1}
\left|\psi_{1,j}^{(0)}\right\rangle &=& 
\sqrt{\frac{2}{N}} \left(\sum_{i=1}^{N}\sin{\left(\frac{\pi}{2}i\right)} c^{+}_{i,j}\right)|0\rangle \, ,
\\
\label{eq:PBC-psi2}
\left|\psi_{2,j}^{(0)}\right\rangle &=& 
\sqrt{\frac{2}{N}} \left(\sum_{i=1}^{N}\cos{\left(\frac{\pi}{2}i\right)} c^{+}_{i,j}\right)|0\rangle \, ,
\end{eqnarray}
and small disorder $W/t'$ lifts the degeneracy without leading to coupling terms. The two states 
\eqref{eq:PBC-psi1} and \eqref{eq:PBC-psi2} are thus the basis of the two-dimensional 
$E^{(0)}=0$ subspace that diagonalizes the corresponding sub-block of the slice Hamiltonian $H_j$ 
with periodic BC when on-site energies taken into account in lowest order. 

The lowest order energy corrections due to the non-zero on-site energies are given by
\begin{eqnarray}\label{eq:PBC-E1}
E_{1,j}^{(1)} &=& 
\frac{2}{N} \left(\epsilon_{1,j}+\epsilon_{3,j}+\epsilon_{5,j}+\dots + \epsilon_{N-1,j}\right)\, ,
\\
\label{eq:PBC-E2}
E_{2,j}^{(1)} &=& 
\frac{2}{N} \left(\epsilon_{2,j}+\epsilon_{4,j}+\epsilon_{6,j}+\dots + \epsilon_{N,j}\right) \, ,
\end{eqnarray}
and the longitudinal hopping terms lead to a coupling strength $t'$ between states 
$\left|\psi_{1(2),j}^{(0)}\right\rangle$ with adjacent values of $j$. In the limit of strong transverse 
coupling $t'/t, W/t \ll 1$, we therefore have at $E=0$ an effective system composed of two uncoupled 
chains with hopping $t'$ and on-slice energies $E_{1/2,j}^{(1)}$ according to 
Eqs.\ \eqref{eq:PBC-E1} and \eqref{eq:PBC-E2}. This readily explains why, in the case 
of periodic BC, the two smallest Lyapunov exponents remain small when $N$ is a multiple of four.  
The Lyapunov exponents for such effective chains are the asymptotic values for the exponents 
$\gamma_1$ and $\gamma_2$ of the quasi-$1D$ strips. Their values at different disorder $W$ 
are shown as green squares on the right vertical axis in Figs.\ \ref{fig:pbc} (b) and \ref{fig:pbc} (d).

The energies $E_{1/2,j}^{(1)}$ are averages over $N/2$ of the 
original on-site energies $\epsilon_{i,j}$, and therefore have a modified probability 
distribution with the reduced variance $\sigma^2 = 2\sigma^2_0/N$. 
A uniform distribution with the effective disorder strength 
\begin{equation}
W_\mathrm{eff}=W\sqrt{\frac{2}{N}}
\end{equation}
has the same reduced variance and leads with \eqref{eq:lyapunov1d} to the approximate value
\begin{equation}\label{eq:lyapunov-eff-pbc}
\gamma_\mathrm{eff}\simeq \frac{0.019}{N} W^2 + O\left[W^4\right]\, .  
\end{equation}
for the Lyapunov exponent of the effective chains.
The dashed line in Fig.\ \ref{fig:gamma_N} (b) represents this result. 

The two smallest Lyapunov exponents of the quasi-$1D$ strips are split by higher order terms of 
the perturbative approach such that the lower one approaches the strong coupling limit 
from below (see Fig.\ \ref{fig:pbc}). The numerical results obtained with $t=100$ and shown as 
green squares in Fig.\ \ref{fig:gamma_N} (b) are therefore slightly below the dashed line representing 
the strong coupling limit \eqref{eq:lyapunov-eff-pbc}. 

\begin{figure*}
\centerline{\includegraphics[width=\linewidth]{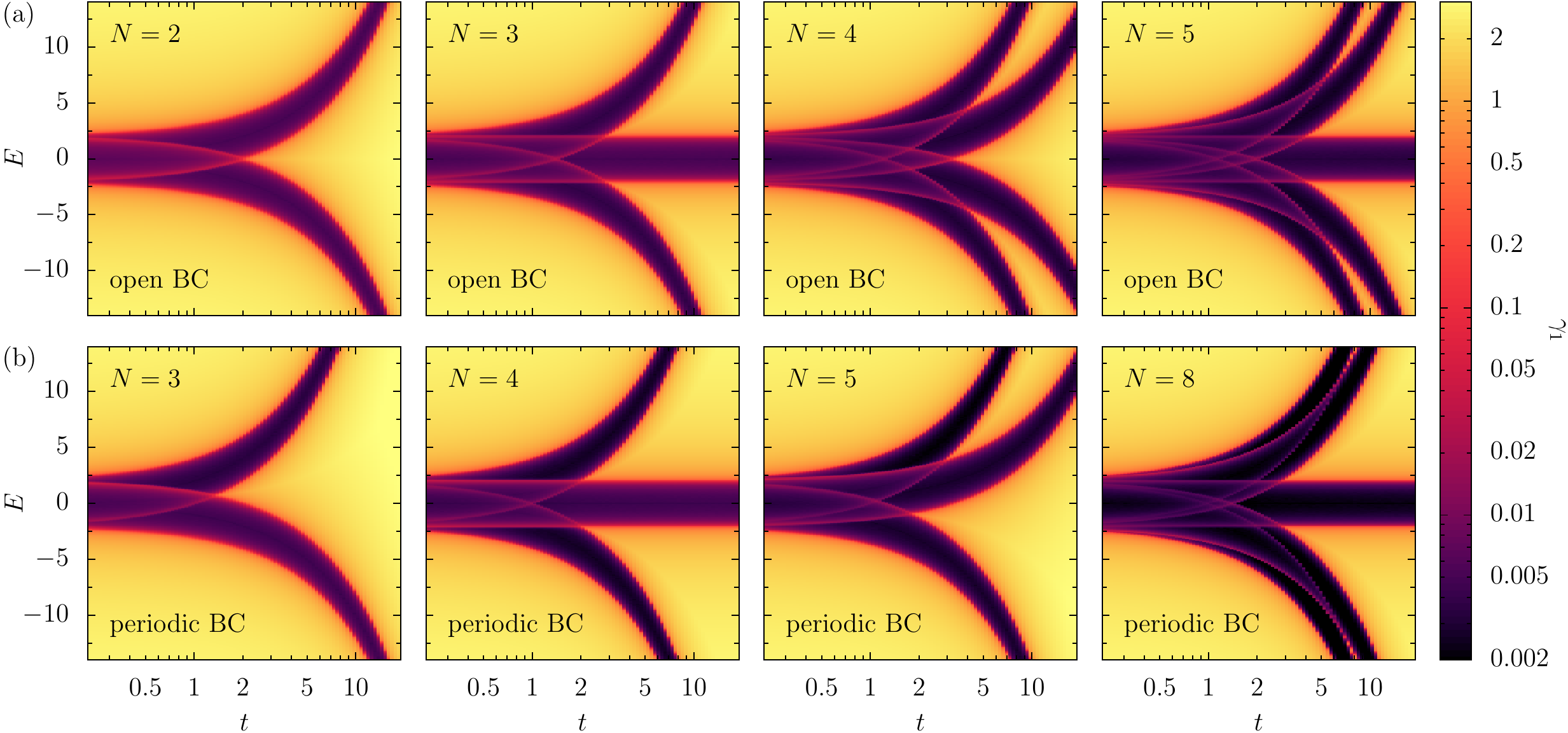}}
\caption{\label{fig:gamma_t_E}
(Color online) The lowest Lyapunov exponent $\gamma_1$, calculated at disorder strength $W=1$, is shown 
in colorscale (grayscale) as a function of the inter-chain hopping $t$ and the energy $E$, with (a) open BC  
and (b) periodic BC, for different values of $N$. It can be observed that energy bands with low $\gamma_1$ 
split at large $t$. 
In the case of  (a) open BC and odd $N$, as well as for  (b) periodic BC and $N$ being a multiple of 4, 
one of the bands is situated around $E=0$, independent of $t$. In all other cases, $E=0$ is located in a 
gap between such bands that becomes wider with increasing $t$.}
\end{figure*}
We have seen for both lateral BC that the quasi-$1D$ strips with strong inter-chain 
coupling $t$ can be mapped onto effective $1D$ chains, as long as $N$ is odd for open BC or a 
multiple of four for periodic BC. Within this constraint, an increase of $N$ increases the number 
of on-site energies  that contribute to the on-slice energies of the transverse slices. 
This gives as a result an effective disorder strength and a corresponding Lyapunov exponent that decrease 
as $W_\mathrm{eff}\propto 1/\sqrt{N}$ and $\gamma_\mathrm{eff}\propto 1/N$, respectively, 
explaining the ``critical" scaling law \cite{20,21,22} of Eq.\ \eqref{eq:scaling}
discussed in Sec.\ \ref{sec:scaling}. 

\section{Discussion}
\label{sec:discussion}

In the previous sections we have presented the combined effects of disorder $W$ and inter-chain 
hopping $t$ in quasi-$1D$ strips which consist of an assembly of $N$ disordered chains. 
Results obtained at $E=0$ are shown in Figs.\ \ref{fig:nbc}, \ref{fig:pbc}, and \ref{fig:gamma_N} 
for a variety of disorders $W$, a very wide range of $t$'s and several values of $N$. 
Our particular focus is on the strong coupling regime $t'/t \to 0$, which is understood via 
a perturbation theory in $t'/t$ and effective chains with diminished disorder 
for special values of $N$. 

In Fig.\ \ref{fig:gamma_t_E} we present colorscale plots of the smallest 
Lyapunov exponent $\gamma_{1}$ as a function of $t$ for $W=1$ and energy $E$ with (a) open 
and (b) periodic lateral BC and several values of $N$. 
A key to our understanding is the evolution of the energy-dependence of $\gamma_{1}$ with 
increasing $t$. At small $t$ all studied systems show similar behavior. The Lyapunov exponents 
are small (dark color) when the energy $E$ lies inside the one-dimensional band $E\in [-2t';2t']$ 
given by the longitudinal hopping $t'=1$. 
The states become more localized in the Lifshitz tails \cite{5} of the spectrum, 
for $2t' < |E| < 2t'+W/2$, whose width is determined by the disorder strength $W=1$. 
For larger absolute values of the energy $|E| > 2t' + W/2$, no electronic states are available, 
the propagation is evanescent and characterized by large Lyapunov exponents (bright color) that 
increase with increasing $|E|$.   

In a clean system ($W=0$) the Hamiltonian is separable in a longitudinal and a transverse part 
with the available total energies being sums of the longitudinal one-dimensional band energy 
and the transverse energy.   
The inter-chain coupling $t$ leads to a discrete spectrum of $N$ transverse energies
with spacings $\propto t/N$ between them that become wider with increasing $t$. 
One can represent the system of $N$ coupled chains in the basis of 
the eigenstates of the clean transverse slices (see Sec.\ \ref{sec:perturbation}) and gets 
$N$ uncoupled channels with energy offsets given by the discrete transverse energies.

This scenario is qualitatively robust against not too strong disorder, 
when the transverse mean free path remains much larger than the width $N$ of the quasi-1D strip. 
In the strong coupling limit one has $t\gg W$, and this condition is  
always fulfilled, at least close to the center of the 1D band, and the basis of the 
transverse channels is the appropriate one for discussing and understanding the 
properties of the quasi-$1D$ strip. 
Moreover, the effective disorder strength of the channels is given by averages of the 
on-site energies as discussed in Sec.\ \ref{sec:perturbation}.
At finite $t$, the disorder breaks the separability of the clean system and couples the 
$N$ channels. While this coupling can be neglected when the channel energies are 
very far from each other, it plays a role at moderate values of $t$ when the bands corresponding to 
neighboring channels overlap, and also in the case of periodic BC where two degenerate channels exist 
independent of $t$.

The spacing of the transverse energies increases beyond the width of the longitudinal band 
(roughly, this happens when $t/N \gtrsim t'$) and the spectrum splits in $N$ subbands that are 
separated by gaps whose width increases linearly with increasing $t$. 
For periodic BC, the number of gaps ($N-1$ gaps) is reduced with respect to the case of open BC 
since some of the transverse states are doubly degenerate. 
For energies situated in one of the subbands, the smallest Lyapunov exponent is small, 
similar to the $1D$ case. In contrast, the gaps of the spectrum are characterized by very large 
Lyapunov exponents, very much like for the energies outside the $1D$ spectrum at small $t$.

In order to understand the zero-energy behavior at strong coupling, the crucial question is whether 
$E=0$ lies in a subband or in a gap. 
The only cases where $E=0$ is inside a subband for all values of the coupling strength $t$ are the 
ones in which the clean transverse problem has a zero eigenenergy. 
As discussed in Sec.\ \ref{sec:perturbation}, this is the case with odd $N$ at open BC and $N$ 
a multiple of four for periodic BC. 
Examples for open BC ($N=3$ and $N=5$) are displayed in Fig.\ \ref{fig:gamma_t_E} (a), and for 
periodic BC ($N=4$ and $N=8$) in Fig.\ \ref{fig:gamma_t_E} (b).
In all other cases shown, all of the subbands increase or decrease in energy proportional to $t$, 
and tend to $\pm\infty$ in the strong coupling limit such that $E=0$ lies in a gap of the spectrum. 
Since the size of the energy gap increases with $t$, the large values of $\gamma_1$ 
observed in these cases further increase with increasing $t$ (see the maxima of the $N$-dependent 
oscillations in Fig.\ \ref{fig:gamma_N}). 
Therefore, the even-odd effect in the number of chains $N$ observed in Fig.\ \ref{fig:gamma_N} (a) 
and the period-of-four oscillations in Fig.\ \ref{fig:gamma_N} (b) are related to $N$-dependent 
oscillations between finite and vanishing density of states at $E=0$.

Related even-odd effects have been found in other systems. The $1D$ to $2D$ crossover is also 
not smooth for the magnetic order in $N$ antiferromagnetically coupled clean spin chains with $S=1/2$. 
While these $2D$ systems exhibit long range order, for even $N$ only short-range magnetic order occurs, 
accompanied by a finite energy gap to magnetic excitations. \cite{23} Also, coupled $d$-wave 
superconducting quantum wires with open BC and half filling have been found to exhibit a parity effect. 
The density of states at $E=0$ is found to be vanishing for even $N$ and finite if $N$ is odd. \cite{24} 
We could also mention Carbon nanotubes which depending on geometry are armchair with an $E=0$ mode 
(metallic) or zigzag without the $E=0$ mode (semiconducting).\cite{22}

In Fig.\ \ref{fig:gamma_t_E} the subband edges are visible at moderate inter-chain coupling 
$t\sim t'=1$ in the form of lines with enhanced $\gamma_1$, even though the subbands are 
overlapping. In this situation, the system corresponds to coupled effective chains 
(one for each subband), one of them being close to the band edge where localization is stronger. 
This is reminiscent of the case of a two-leg ladder composed of two coupled chains having different 
localization length studied in Ref.\ \onlinecite{17}, where a similar enhancement of the 
localization strength was found for energies close to the band edge of the strongly localized chain.
The crossings of the band edges as a function of $t$ with $E=0$ (a horizontal line in 
Fig.\ \ref{fig:gamma_t_E}) are the origin of the peak 
structure observed in Figs.\ \ref{fig:nbc} and \ref{fig:pbc} at $t\gtrsim 1$. 
The number of peaks increases with the number of non-degenerate subbands that deviate from 
$E=0$ at strong coupling. 

\section{Conclusions}
\label{sec:conclusions}

We have studied the dimensionality crossover from $1D$ to $2D$ for $N$ coupled chains 
with disorder $W$ and inter-chain coupling $t$ as $N$ increases. In $2D$, the lower critical 
dimension for Anderson localization, all states are localized by disorder unless time reversal and 
spin-rotation symmetry is broken. We find no smooth crossover from $1D$ to $2D$ as a function of $N$, 
but parity-dependent Anderson localization in the presence of disorder $W$ and strong 
inter-chain coupling $t$.

Our main result is an unexpected effect of the parity of $N$ on the behavior of the smallest 
Lyapunov exponent $\gamma_{1}$ at $E=0$. An even-odd effect for open BC and a multiple-non-multiple 
of four effect for periodic BC is shown in Figs.\ \ref{fig:nbc}, \ref{fig:pbc}, and \ref{fig:gamma_N}.
This parity effect implies ``immunity" to the strong localization obtained for large $t$, for even $N$ 
with open BC and $N$ non-multiple of four for periodic lateral BC. The strong inter-chain hopping $t$ 
reduces the strength of localization even below the weakly coupled (small $t$) case for some $N$'s, 
while  for other $N$'s localization for large $t$ is much stronger than in $1D$. 
The weaker Anderson localization for large $t$ for some $N$'s and the gaps in the spectrum which 
lead to stronger localization for other $N$'s are quantitatively explained via a perturbative treatment 
in the strong inter-chain coupling limit $t'/t, W/t \to 0$, where the system can be mapped onto an 
effective model of one (two) weakly disordered chain(s) arising from the one (two) zero-energy states 
in the spectrum of clean transverse slices with open (periodic) BC. 
Our treatment also explains the intermediate critical scaling of Eq.\ \eqref{eq:scaling} found in 
many disordered systems.

The $E=0$ state studied usually has the largest localization length. Similar results are obtained for 
other energies within the band of a clean $1D$ chain $[-2t';2t']$. The parity of $N$ effect can 
have consequences for finite size scaling studies where results for $N\to \infty$ are obtained 
from rather small $N$'s. In our case $\gamma_1$ does not depend smoothly on $N$ as required.   
The effect is related to topological\cite{25,26} ones, and the integer $N$ is like a winding 
property which affects Anderson localization. 

This work was partially motivated by recent experiments on optical wave guide arrays. \cite{10,11,12,13} 
In these works light propagates along the waveguides in $z$-direction and Anderson localization in 
the transverse $x$--$y$ plane is studied experimentally and theoretically by investigating the 
spreading of a local excitation to neighboring waveguides. 
Anisotropy in the couplings is introduced via different mean distances in $x$ and $y$-direction, 
and a randomization of the distances in one direction introduces off-diagonal disorder.
In Ref.\ \onlinecite{10}, Anderson localization was shown to weaken by increasing $N$, 
hence in going from $1D$ to $2D$. However, the localization length is of the order of a 
lattice spacing. We predict that in the case of strong anisotropy $t/t'$ the 
parity of $N$ should play an important role, provided the disorder is weak and the 
localization length larger than the lateral size of the system. 
Then, at odd $N$ the system should remain weakly localized for larger inter-chain coupling 
as it does for small $t$. 

In summary, our study shows that Anderson localization for states close to $E=0$, in a disordered 
quasi-$1D$ system of $N$ chains coupled by inter-chain hopping $t$, depends dramatically on the value 
of $N$. For small $t$ localization becomes weaker on going from $1D$ to $2D$ (increasing $N$) while 
for large $t$ localization becomes stronger for some $N$ and weaker for other $N$. 
The reduced localization for large $t$ arises when the transverse energy splitting exceeds the width of 
the longitudinal $1D$ subbands. In conclusion, the interplay between disorder $W$ (causes localization) 
and strong anisotropy $t$ (creates gaps) is shown to depend on the number of chains $N$. Only at very 
large $N$ a smooth crossover from $1D$ to $2D$ is reached. The increase of $t$ requires higher $N$ to 
suppress the $N$-dependent oscillations in the localization.

\end{document}